\documentclass{optica-article}
\journal{opticajournal} 
\usepackage{bm}
\usepackage{
    lineno,
    textcomp, 
    float,
    mathrsfs
    }
\articletype{Research Article}

\usepackage{booktabs}
\usepackage{hhline} 
\usepackage{tabularray}

\newcommand{\vect}[1]{\mathbf{#1}}
\newcommand{\vectrho}{\bm{\rho}}

\newcommand{\secref}[1]{Sec.~\ref{#1}}
\newcommand{\eqnref}[1]{Eq.~(\ref{#1})}
\newcommand{\figref}[1]{Fig.~\ref{#1}}

\newcommand{\maxwell}[0]{\text{s.t. }\ \nabla\times\frac{1}{\mu}\nabla\times\vect{E}-\omega_m^2\vect{\varepsilon}(\vectrho)\vect{E}=\ -j\omega_m\vect{J}}

\begin{document}


\title{Integrated Photonic Topology Optimization with Nonvertical Sidewall Profiles: Applications in Lithium Niobate and Silicon}

\author{Michael~J.~Probst\authormark{1}, Jacob~M.~Hiesener\authormark{1}, Archana~Kaushalram\authormark{1}, and Stephen~E.~Ralph\authormark{1,*}}

\address{\authormark{1}School of Electrical and Computer Engineering, Georgia Institute of Technology, Atlanta, GA 30308, USA\\
\email{\authormark{*}stephen.ralph@ece.gatech.edu}} 


\begin{abstract*} 
We enable density-based topology optimization (TO) to design integrated photonic devices featuring nonvertical sidewall profiles. Specifically, we demonstrate TO for fabrication processes with slanted sidewalls which are often used to enhance vertical coupling efficiency and fabrication processes with angled sidewalls which are a common feature of etching. The techniques demonstrated are readily adaptable to other etch profiles such as asymmetric or nonlinear. The enhancements are compatible with existing TO techniques, lengthscale constraints and multi-layer designs, and any dielectric materials, suiting the techniques for both academic and commercial foundry fabrications. We demonstrated the developed capabilities by designing slanted and angled silicon grating couplers and thin-film lithium niobate on insulator dual-polarization s-bends. 
\end{abstract*}
\section{Introduction}

\begin{figure}[ht!]
    \centering
    \includegraphics[width=\textwidth]{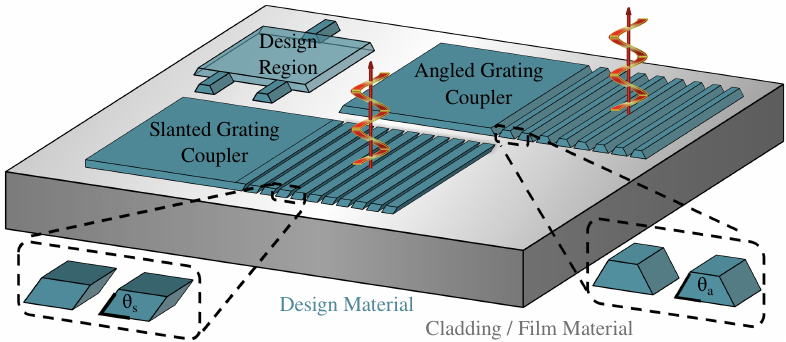}
    \caption{Integrated photonic structures with angled and slanted etch profiles. Angled profiles arise from the inherent difficulty in achieving perfectly vertical planes during the etching process (\secref{sec:angle}). In contrast, slanted etch profiles are a deliberate design choice aimed at improving out-of-plane coupling, while being more difficult to fabricate (\secref{sec:slant}). These profiles are observed across a range of material platforms such as silicon (Sec. 3,4) and lithium niobate (\secref{sec:sbend}) are must be considered with in TO context. In this work, angled sidewalls are denoted with $\theta_a$ and slanted sidewalls with $\theta_s$.}
    \label{fig:grating_motivation}
\end{figure}

Photonic topology optimization (TO) \cite{sigmund_overview,hammond_phase,Molesky2018,shang2023inverse,de2024topology,deng2024inverse,carbide_inverse} is a sophisticated inverse-design approach to produce optical devices by parameterizing a design region into millions of voxels, and optimizing these voxels as unique degrees of freedom.  This approach results in novel freeform designs characterized by nonintuitive geometries, exceptionally compact sizes, and unparalleled performances. Maturing fabrication processes facilitate increasingly complex geometries that necessitate enhancements to conventional TO techniques in order to optimize these geometries. For example, etching slanted sidewalls is a technique to improve directionality in out-of-plane couplers (\figref{fig:grating_motivation}). Angled sidewalls, on the other hand, are often an undesired artifact of dry etching materials such as silicon and lithium niobate in both foundry \cite{lee2024low} and laboratory \cite{LNLoncar} settings. However, tuning the sidewall profile by varying the etch chemistry of the dry etch process provides another degree of freedom in the design process \cite{sidewallTuning}. Accurately simulating and optimizing slanted and angled geometries is necessary to realistically assess performance prior to device fabrication, leading to improved performance and first-time-right design \cite{shang2023inverse}.

Optimizing geometries with nonvertical etch profiles in a density-based TO setting requires carefully choosing projection techniques that map planar design variables into the vertical direction. Any operations used must be compatible with grayscale geometries (as the design variables can take on intermediate values between the material permittivites) and be differentiable. As such, the two principal operations used in this work, the shifted-$\delta$---a filter used to constructed slanted geometries (\secref{sec:slant})---and the $\mathcal{S}$Morph \cite{kirszenberg2021going}---an erosion/dilation operation used to construct angled sidewall geometries (Sec. 4,5)---both satisfy these requirements and so were chosen among other candidates. By developing novel TO compatible parameterizations for increasingly complex geometries, we broaden the class of design problems solvable by TO.



\subsection{Slanted sidewalls} 
Etch profiles featuring slanted sidewalls serve as useful building blocks for intricate three-dimensional microstructures which are pivotal in a myriad of applications such as grating couplers, photonic crystals, and nanopillar arrays \cite{slantEtch}.  Slanted grating couplers operate with high coupling efficiency, outperforming their vertical counterparts for the following reasons. First, their asymmetric slanted parallelogramic profile breaks the vertical symmetry inherent in surface normal coupling between fiber and the planar waveguide, which enables coupling of light exclusively into the desired waveguide direction while suppressing coupling in the opposite direction \cite{slantedGC1}. Moreover, parallelogramic shaped gratings also generally have a larger radiation factor and higher radiation directionality compared to other grating profiles \cite{slantedGC2}. Slanted grating couplers are also crucial in X-ray imaging and in diffractive optical elements used to couple a projected image into a waveguide in displays for augmented and virtual reality glasses \cite{Xray,ARVR}. Another instance of slanted etching is to create oblique or slanted nanopillars in the surface texturing of silicon-based photovoltaic devices that has been shown to greatly improve antireflection properties \cite{obliquenanopillars} and improve surface-enhanced Raman scattering of the silicon substrate \cite{BlackSi}. Compared to conventional vertical nanopillar arrays, these structures also provide superior electrical contact, potentially enhancing the power conversion efficiency of solar cells \cite{solarcell}. Optical reflectance is significantly reduced when employing oblique pillars due to the elongated paths for light trapping created by the trenches between these structures. Therefore, slanted silicon nanowires exhibit an improved absorption compared to vertical nanowires for a large range of incident angles over infrared, visible, and ultraviolet regions of the solar spectrum which is crucial for solar cells \cite{slantednanoWire}. Nonuntivitve, high performing inverse designed structures, combined with the inherent advantages of slanted etch geometries\cite{probstslanted} may enable record breaking devices in efficiency, directionality, and performance for a variety of applications.


\subsection{Angled sidewalls}
Etching perfectly vertical sidewalls in certain photonic processes is challenging, so etched materials often feature a sloped etch, called angled sidewalls. These angled sidewalls are typically an undesirable artifact of the dry etching chemistry and often degrade device performance; however, their effects can be mitigated or potentially rendered useful during the design phase through proper simulation and optimization techniques. Furthermore, in some materials, the angled sidewalls can be tuned by modifying the dry etch chemistry \cite{sidewallTuning} to achieve positive or negative slope of the waveguide sidewalls \cite{sidewall1,sidewall2}. The sidewall angle is an additional degree of freedom which has been exploited to design mode converting devices that are shorter than could be achieved with vertical sidewalls \cite{modeHybrid1,modeHybrid2} and to design slot waveguides with enhanced nonlinearity in the mode \cite{slot}.


Angled sidewalls are found in both commercial foundry silicon-on-insulator (SOI) platforms and specialty material platforms such as litium niobate on insulator (LNOI) \cite{lu2024two} and lithium tantalate \cite{Powell:24}. The most common dry etching for LNOI is pure physical etching with ArC plasma or ArC milling, which has a low etch selectivity between lithium niobate and common hard mask materials such as SiO$_2$ and amorphous silicon. The low etch selectivity restricts the etch depth leading to shallow ridge waveguides while the mechanical nature of the etch results in a trapezoidal cross section with the sidewall angles ranging from $32^\circ$ to $80^\circ$ \cite{LNLoncar,slantedGC3,slantedGC5,slantedGC4}.  Recent efforts try to overcome this challenge using diamond-like-carbon as a hard mask process and reports a fully etched strip waveguide with the sidewall angle of $80^\circ$ \cite{diamondLN}. Angled sidewalls are also observed in densely packed nanopillars in III-V compound semiconductors such as InGaAs \cite{InGaAs}. As the integrated photonics ecosystem continues to push the fabrication limits toward smaller feature sizes and more complex geometries, as well as explore novel material platforms, intelligent algorithms crafted to optimize these structures are essential to leverage the full benefit of modern fabrication techniques.
 
\subsection{Overview}
The paper is organized as follows: in \secref{sec:TO} we provide an overview of the TO formulations used in this work. In \secref{sec:slant}, we develop TO for slanted geometries, and demonstarted the capabilities by designing 1D SOI grating couplers with varying slant angle and feature size. In \secref{sec:angle}, we formulate TO for topologies with angled sidewalls, again demonstrated with 1D SOI grating couplers with varied sidewall angles. Lastly, in \secref{sec:sbend}, we present a more complex optimization: a 3D dual-polarization s-bend on z-cut thin-film LNOI. The s-bend outperforms a Bezier curve design in all metrics and showcases the capabilities of TO for angled sidewalls.


\section{Photonic topology optimization overview}\label{sec:TO}
In this section, we summarize our previous efforts in density-based photonic TO \cite{probst2024fabrication,hammond_constraints,hammond2022high} and detail the specific formulations used in this work. We optimize $N$ objective functions $(f_1,...,f_N)$ as a function of Maxwell's equations at $M$ frequency points on a geometry subject to $K$ constraint functions $(g_1,...,g_K)$. Specifically:
\begin{equation}\label{eq:obj}
    \begin{matrix}
         \min_{\vectrho}\left[\displaystyle\sum\limits_{n=1}^N\bar{g}_n\big(f_n(\vect{E}),\vect{q}_n\big)+\sum\limits_{k=1}^K\bar{g}_{N+k}\big(g_k(\vectrho),\vect{q}_{N+k}\big)\right] & n\in\left\{1,...,N\right\},\ k\in\{1,...,K\}\\
        \maxwell & m\in\left\{1,2,...,M\right\}\\
        0\leq\vectrho\leq1&\
    \end{matrix}
\end{equation}
where $\bar{g}_i,\ i\in\{1,...,N+K\}$ are differentiable spline-based scaling functions \cite{messac1996physical} that condition the optimization problem, $\vect{q}$ are user-defined lists that define the scaling of each objective/constraint, and $\vect{E}$ and $\vect{J}$ are the electric field and source current respectively. The device permittivity $\varepsilon$ is parameterized from latent (unfiltered) design variables $\vectrho$ using a filter-threshold parameterization scheme\cite{sigmund_overview}. In the filter step, the latent 2D (1D) variables are convolved with a conic (triangular) filter to produce the filtered design variables $\widetilde\vectrho$:
\begin{equation}\label{eq:filter}
    \begin{matrix}
        \widetilde{\vectrho}=w(\vect{x})*\vectrho && w(\vect{x})=
    \begin{cases} 
      \frac{1}{a}\left(1-\frac{|\vect{x} - \vect{x}_0|}{R}\right) & \vect{x} \in \mathcal{N} \\
        \quad 0 & \vect{x} \notin \mathcal{N}
   \end{cases}
    \end{matrix}.
\end{equation}
$R$ is the radius of the filter which encourages a minimum feature size in the geometry and $a$ is a normalization factor such that $\smallint d\vect{x} \, w(\vect{x})=1$. Next, the parameters are encouraged toward binarization using a modified hyperbolic tangent projection:
\begin{equation}\label{eq:threshold}
    \bar{\vectrho}=\frac{\rm tanh \left(\beta\eta\right)+\rm tanh \left(\beta\left(\vect{{\widetilde{\vectrho}}}-\eta\right)\right)}{\rm tanh\left(\beta\eta\right)+\rm tanh\left(\beta\left(1-\eta\right)\right)},
\end{equation}
where $\beta$ and $\eta$ control the steepness and center of the projection respectively. Lastly, the electric permittivity is linear interpolated from the thresholded design variables using
\begin{equation}\label{eq:permittivity}
    \vect{\varepsilon}_r(\bar{\vectrho})=\vect{\varepsilon}_\text{min}+\bar\vectrho\big(\vect{\varepsilon}_\text{max}-\vect{\varepsilon}_\text{min}\big).
\end{equation}
The open-source finite difference time domain (FDTD) simulator Meep \cite{meep} is used to perform the forward and adjoint simulations. Then, the device performance and gradient of the FOM with respect to all design variables are evaluated for all design wavelengths simultaneously using the hybrid time-frequency adjoint variable method from just two Maxwell solves \cite{Strang07,steven_adjoint,hammond2022high}. We backpropagate the gradient through the filter-thershold parameterization with a vector-Jacobian product (vJp)\cite{baydin2018automatic} using the open-source package Autograd\cite{maclaurin2015autograd}.  The latent design variables, $\vectrho$, are then optimized using the gradient information and the globally convergent method of moving asymptotes (GCMMA)\cite{svanberg_mma}, provided by the open-source nonlinear optimization package NLopt\cite{nlopt}.

\section{Slanted sidewalls}\label{sec:slant}
\begin{figure}[ht!]
    \centering
    \includegraphics{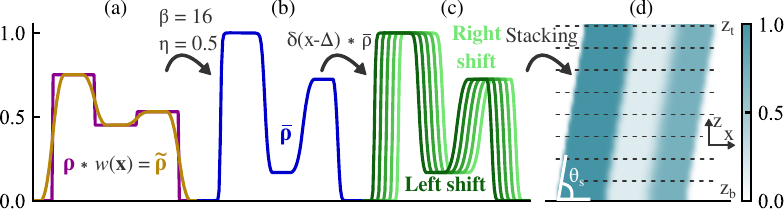}
    \caption{The TO parameterization scheme tailored for slanted sidewalls. The latent design variables $\vectrho$ (a) are filtered with a triangular filter $w$ (\eqnref{eq:filter}) and (b) thresholded with the modified hyperbolic tangent function (\eqnref{eq:threshold}) with center and steepness $\eta$ and $\beta$. (c) The thresholded parameters are (d) vertically stacked after undergoing a shift dependent on the height in the layer and the user-defined slant angle. The shift is realized by convolving the design variables with a shifted $\delta$-function, which is compatible with the denisty-based TO design flow.}
    \label{fig:slanted_mapping}
\end{figure}

Slanted etch profiles yield excellent out-of-plane coupling with just a single design layer by reducing substrate emissions and backreflections. Their complicated geometries, albeit difficult to design analytically, are easily handled by TO frameworks. To realize a slanted etch angle on a device, the design variables are copied (\figref{fig:slanted_mapping}c) and vertically stacked  (\figref{fig:slanted_mapping}d) after undergoing shift dependent on the height in the layer and the user-defined slant angle, $\theta_s$. The shift is implemented by convolving the thresholded design variables, $\bar\vectrho$ with a shifted $\delta$-function given by:
\begin{equation}
    \begin{matrix}
        \bar\vectrho' = \bar\vectrho \ast \delta(x - \Delta) && 
        \Delta = \cfrac{z-(z_t-z_b)/2}{\tan(\theta_s)}    
    \end{matrix},
\end{equation}
where $\Delta$ is the amount of shift applied to each copy of the design variables, $z$ is the height in the layerwhere the copy will be placed, and $z_t$ and $z_b$ are the top and bottom height of the layer respectively (\figref{fig:slanted_mapping}). The shifted-$\delta$ convolution preserves the differentiability of the projection while simultaneously being compatible with both grayscale and binary geometries. This copying and stacking process is extensible to 2D and 3D geometries which and can produce a slant angle of any magnitude in any direction. Note that for 2D images of design variables, the slant angle can point in any direction. Therefore, the thresholded design variables are convolved with a $\delta$-function of both transverse directions (i.e. $\delta(x-\Delta,y-\Gamma)$ where $\Gamma$ is the shift in the $y$-direction).

We illustrated the design process by optimizing 1D silicon grating couplers with varying slant angles. 1D grating couplers can efficiently couple to standard single mode fiber (SMF) because the mode of a 12 $\upmu$m waveguide has 97\% overlap in the transverse direction with the fiber mode \cite{taillaert2004compact,probst2024inverse}. Furthermore, the effective index in a 12 $\upmu$m waveguide ($n_\text{eff}=2.8302$) is nearly identical to that of a SOI slab mode ($n_\text{eff}=2.8309$) and can therefore be simulated as a slab mode without loss of accuracy \cite{taillaert2004compact}. These two assumptions simplify grating design to just two dimensions (\figref{fig:grating_motivation}) and still enable excellent experimental performance. Furthermore, the computational cost is greatly reduced allowing for rapid design to explore the effects of slant angle.

The gratings were optimized according to \eqnref{eq:obj} with an FOM at each design wavelength given by:
\begin{equation}
    \label{eq:FOM}
    f_n(\vect{E})=10\times\log_{10}\left(\cfrac{|\alpha_0^+|^2}{P_n}\right),
\end{equation}
where $\alpha_0^+$ is the overlap of the simulated fields excited from the fiber mode and the fundamental mode of the integrated waveguide pointed away from the grating and $P_n$ is power emitted by the source which normalizes the overlap integral. The mode overlap for the $m^\text{th}$ mode (in either propagation direction) is given by:
\begin{equation}
    \label{eq:overlap}
    \alpha_m^\pm=\int\limits_A\left[\vect{E}^*(r)\times \vect{H}_m^\pm(r)+\vect{E}_m^\pm\times\vect{H}^*(r)\right]\cdot d\vect{A},
\end{equation}
where ``$+(-)$'' denotes the forward (backward) propagating modes, $d\vect{A}$ is the differential cross-sectional area of the waveguide pointing along the direction of propagation, $\vect{E}(r)$ and $\vect{H}(r)$ are the amplitudes of the simulated electric and magnetic fields at optimization wavelengths, and $\vect{E}^\pm_m(r)$ and $\vect{H}^\pm_m(r)$ are the mode profiles for the $m^\text{th}$ integrated waveguide mode. For the current optimization problem, we defined one FOM at each $\lambda=1540\text{~nm, }1550\text{~nm, } 1560$~nm. The gratings were optimized using a $10\ \upmu$m long design region to couple the TE$_0$ integrated waveguide mode into SMF positioned $1\ \upmu$m above the silicon with a launch angle of $15^\circ$. The SMF mode was modeled by a Gaussian source with $10.4\ \upmu$m beam waist\cite{marchetti2017high}. The simulations were performed at of resolution $100\ \text{voxels}/\upmu$m, and the optimization was performed with a resolution of $200\ \text{voxels}/\upmu$m\cite{hammond2022high}. The higher optimization resolution is enabled because Meep interpolates between the Yee grid and design grid \cite{hammond2022high}.

\begin{figure}[ht!]
    \centering
    \includegraphics[width=\textwidth]{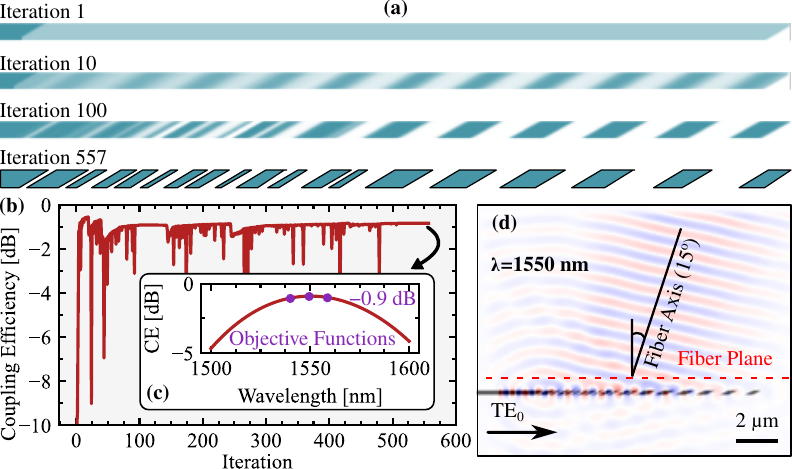}
    \caption{(a) Design evolution of a slanted grating coupler with material slant angle of 30$^\circ$ and minimum feature size of 80~nm. The slant angle is present throughout the entire optimization, with iterations 1, 10, 100, and 557 shown. (b) The maximum coupling efficiency of the grating during the optimization and (c) final performance of the completely binarized design. (d) A field plot of the grating excited by waveguide mode at $\lambda=1550$~nm showing strong overlap with the fundamental fiber mode at launch angle $15^\circ$ and minimimal substrate (downward) emission due to the nonvertical etch profile.}
    \label{fig:slanted_evolution}
\end{figure}

Slanted grating couplers in silicon have been demonstrated with slant angle between $32^\circ$ to $80^\circ$ degrees using direct etching with a focused ion beam \cite{slantedGC3,slantedGC5,slantedGC4} or plasma etching techniques \cite{slantEtch}. \figref{fig:slanted_evolution} exemplifies the design process for a grating with 30$^\circ$ slant angle and 80~nm feature size. Beginning with a grayscale initialization, the optimizer readily found a grayscale grating structure using the slanted sidewalls (\figref{fig:slanted_evolution}a). Enforcing the minimum feature size and binarization initially caused a reduction in the performance of the device as shown in the performance plot (\figref{fig:slanted_evolution}b), but the final feature size compliant design had excellent broadband performance (\figref{fig:slanted_evolution}c) of $-0.9$~dB insertion loss and 93~nm 3-dB bandwidth. The field plot of the grating excited from the waveguide with $\lambda=1550$~nm  (\figref{fig:slanted_evolution}d) shows strong coupling upward at the desired launch angle of 15$^\circ$ and minimal power emitted toward the substrate and higher order modes. Because the devices are optimized using a 1D vector of design variables, we do not see the nonintuitive freeform structures typical of inverse-designed devices. However, because the transverse mode profile of the waveguide is already closely matched to that of the SMF\cite{taillaert2004compact}, we would not expect significant curvature or apodization in the grating teeth\cite{HammondGrating} if the device were designed using the full 3D solver.

\begin{figure}
    \centering
    \includegraphics[width=\textwidth]{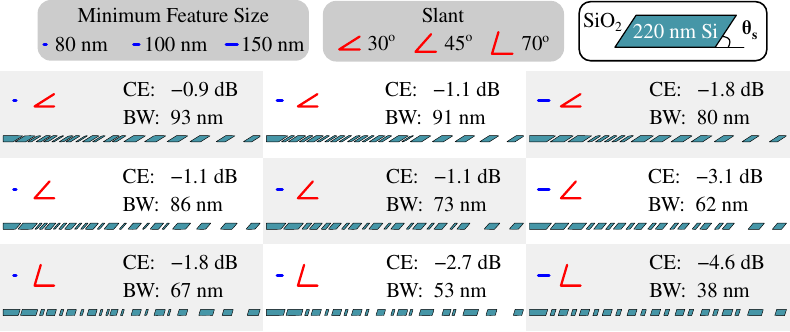}
    \caption{A study of grating couplers designed in 220~nm SOI with slant angles $\theta_s$ of 30$^\circ$, 45$^\circ$, and 70$^\circ$ and feature sizes of 80~nm, 100~nm, and 150~nm. Sharper etch angles and smaller feature sizes unequivocally yield higher coupling efficiencies and broader 3-dB bandwidths. The geometry with $\theta_s=30^\circ$ and minimum feature size of 80~nm had the best coupling of $-0.9$~dB and 93~nm bandwidth while the geometry with $\theta_s=70^\circ$ and minimum feature size of 150~nm had the lowest coupling of $-4.6$~dB and 38~nm bandwidth.}
    \label{fig:slanted_study}
\end{figure}

Next, we show the effects of slant angle and feature size on final coupling efficiency by designing 1D grating couplers at 30$^\circ$, 45$^\circ$, and 70$^\circ$ (\figref{fig:slanted_study}). Each slant angle was designed with three different mask sizes: 80~nm, 100~nm, and 150~nm for a total of nine designs. The gratings designed with different slant angles and feature sizes showed a clear trend: sharper slant angles and smaller feature sizes yielded both higher coupling efficiencies and broader 3-dB bandwidths. Since the minimum feature size is imposed on the width of the mask, fabricating a structure with slant angle geometry necessitates a tool that can etch finer than the mask size. Due to the conservative feature sizes imposed, this is still possible with modern etching techniques\cite{gao2020plasmonic,harinarayana2021two,albrechtsen2022nanometer,pang2021inverse}.   

We note the shifted-$\delta$ implementation of the slanted profile is automatically compatible with other aspects of TO such as minimum feature size and minimum area constraints \cite{hammond_constraints,zhou_geometric}. Since the design variables are not modified as they are stacked in the layer, the mask size and the feature size are the same. Therefore, the minimum feature size can be enforced on the unshifted design parameters same as for TO with vertical sidewalls.


\section{Topology optimization with angled sidewalls}\label{sec:angle}
\begin{figure}[ht!]
    \centering
    \includegraphics{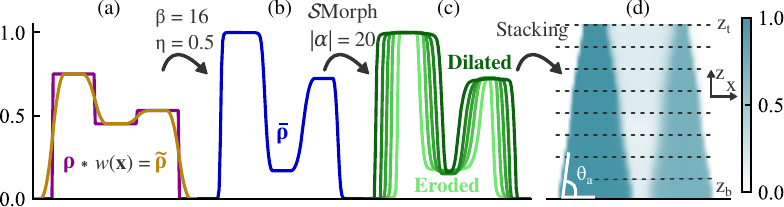}
    \caption{The TO parameterization scheme with angled sidewalls. (a) The latent design variables $\vectrho$ are filtered with a conic filter $w$ (\eqnref{eq:filter}) to generate the filtered design variables $\widetilde\vectrho$ and (b) thresholded with the modified hyperbolic tangent function with center and steepness parameters $\eta$ and $\beta$ (\eqnref{eq:threshold}). (c) The thresholded parameters undergo a height-dependent erosion or dilation from the $\mathcal{S}$Morph operation at every voxel height in the layer and are (d) vertically stacked to make geometries with angled etch profiles.}
    \label{fig:angled_mapping}
\end{figure}

We present a novel parameterization technique that enables density-based TO for devices that are etched with angled sidewalls. Including angled sidewalls in a TO design flow introduces additional complications and must be performed in a manner compatible with all aspects of TO. The design parameters must be nonuniformly extruded into the vertical dimension to construct a geometry with angled sidewalls. Any transforms used to enact the nonvertical extrusion must be differentiable and compatible with grayscale design parameters used in density-based TO. The nonuniform extrusion is accomplished by a height-dependent morphological transform applied to the thresholded design variables (\figref{fig:angled_mapping}a-c), and vertically stacking these transformed parameters (\figref{fig:angled_mapping}d) to produce a user-defined sidewall angle. We use the $\mathcal{S}$Morph transform\cite{kirszenberg2021going}, which is a differentiable, grayscale-compatible erosion/dilation defined as:
\begin{equation}\label{eq:smorph}
    \mathcal{S}\text{Morph}(f,w,\alpha)(x)=\frac{\displaystyle\sum\limits_{y\in W(x)}\big(f(y)+w(x-y)\big)e^{\alpha(f(y)+w(x-y))}}{\displaystyle\sum\limits_{y\in W(x)}e^{\alpha(f(y)+w(x-y))}},
\end{equation}
where $\alpha$ is a user defined parameter that defines the smoothness of the projection and $w$ is a structuring element that determines the amount of erosion or dilation. The $\mathcal{S}$Morph operation accomodates pseudo-dilation $(\alpha>0)$ and pseudo-erosion $(\alpha<0)$, and as $|\alpha|$ becomes large, the $\mathcal{S}$Morph operation converges to the non-differentiable grayscale erosion or dilation operation. For this work, different values of $\alpha$ were used during the optimization, and all final simulations presented used $|\alpha|=30$ which is sufficiently close to the nondifferentiable dilation/erosion \cite{kirszenberg2021going}. The structuring element for the transform is defined by the height within the layer and the desired sidewall angle. In this work, we used a line segment for 1D erosions/dilations and a circle for 2D erosions/dilations with the radius, $L$, given by:
\begin{equation}\label{eq:erosion}
    L=\cfrac{z-(z_t-z_b)/2}{\tan(\theta_a)},
\end{equation}
where $z$ is the height in the layer, $z_t$ and $z_b$ are the top and bottom heights of the layer respectively, and $\theta_a$ is the sidewall angle (\figref{fig:angled_mapping}d). Note that the thresholded design parameters, without undergoing any morphological transform, represent the design parameters at the midpoint height of the layer. See Appendix A,B for alternate erosion/dilation methods that satisfy some but not all of the design requirements.



Since $\mathcal{S}$Morph operation is performed after the filter-threshold operations, there are additional considerations concerning minimum feature size enforcement. We utilize lengthscale constraints \cite{zhou_geometric,hammond_constraints} to enforce the minimum feature size which depend on the interplay of the unfiltered, filtered, and thresholded design variables. The thresholded design variables, as mentioned above, represent the geometry at the midheight of the layer (i.e. $z=(z_b+z_t)/2$) and therefore do not undergo the morphological transform because the design variables are eroded above the center and dilated below the center. However, during fabrication, the minimum feature size is determined by the fabrication mask, which corresponds to the design variables at the \textit{top} of the design layer. Therefore, we must enforce a larger minimum feature size on the center of the design layer, specifically, the filter radius $R$ (\eqnref{eq:filter}) is given by: 
\begin{equation}\label{eq:feature}
    R=\cfrac{1}{2}\left(f_{\text{mask}}+\cfrac{z_b-z_t}{|\tan(\theta_a)|}\right)
\end{equation}
where $f_{\text{mask}}$ is the minimum feature size of the mask. We also enforce the same feature size for the minimum linespacing as the linewidth, which results in a linespacing at the bottom of the layer equal to the mask width. 

\begin{figure}[ht!]
    \centering
    \includegraphics[width=\textwidth]{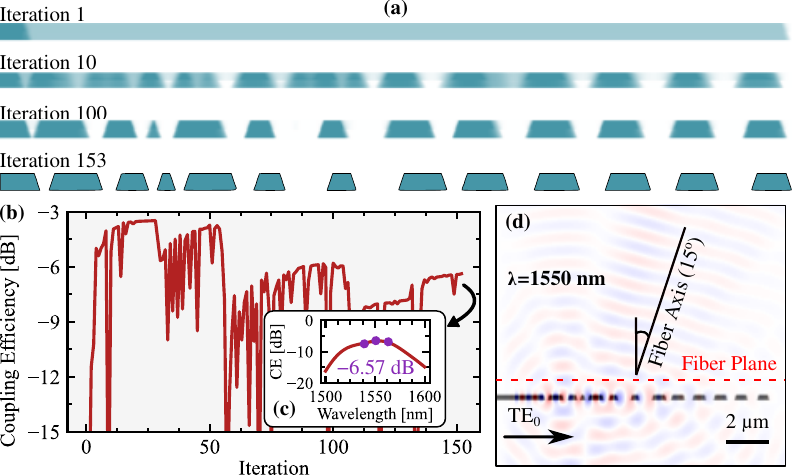}
    \caption{(a) Design evolution of a SOI grating coupler with 70$^\circ$ angled sidewalls. The gray initialization, iterations 10 and 100, and the final binarized design are shown. During optimization, the optimizer quickly found a high performing grayscale design (b) and maintained that performance during gradual binarization and minimum feature size adherence. (c) The grating achieved $-6.57$~dB coupling efficiency with a binary design. (d) When excited from fundamental eigenmode source, the grating emits at a $15^\circ$ launch angle; however, a significant amount of power is lost as leakage downward and into higher order diffraction orders.}
    \label{fig:angled_evolution}
\end{figure}

We illustrated the optimization process for geometries with angled sidewalls by optimizing 1D SOI grating couplers with varying sidewall angles. The optimization problem, layer thicknesses, resolutions, and FOM are identical to the gratings optimized in \secref{sec:slant}.  \figref{fig:angled_evolution} shows the design evolution of an angled grating with $70^\circ$ etch angle and 80~nm feature size. The 70$^\circ$ angle was present in the design geometry throughout the optimization, and the geometry completed binarization after 153 iterations with a final performance of $-6.57$~dB (\figref{fig:angled_evolution}a-c). 

\begin{figure}
    \centering
    \includegraphics[width=\textwidth]{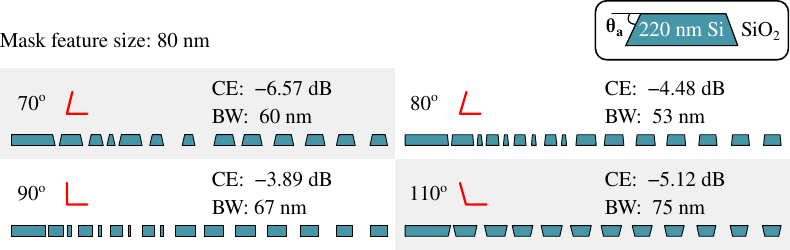}
    \caption{A study of 1D SOI fiber-chip grating couplers designed over four etch angles (70$^\circ$, 80$^\circ$, 90$^\circ$, and 110$^\circ$) and mask feature size of 80~nm. The gratings were clad in SiO$_2$, so vertical symmetry is only broken by the nonvertical etch profile. The vertical geometry exhibited the best performance, likely due to the smaller feature size than the geometries with nonvertical etch profiles. The undercut geometry underperformed the gratings with 80$^\circ$ and 90$^\circ$ etch angles but had higher coupling than the grating with 70$^\circ$ etch angle.}
    \label{fig:angled_study}
\end{figure}

\figref{fig:angled_study} shows the geometries of gratings optimized over four sidewall angles: 70$^\circ$, 80$^\circ$, 90$^\circ$, and 110$^\circ$, which represent a relatively shallow etch angle, a steep etch angle, a perfectly vertical etch, and an undercut etch angle in silicon, respectively. All devices were designed with a mask minimum feature size of 80~nm, a typical minimum feature size for integrated silicon photonics. While the nonvertical etch profile breaks z-symmetry, theoretically enabling unity coupling efficiency, the shallow etch and undercut etch gratings had lower coupling efficiency than the vertical etch. We identify two reasons for this: \textit{i}) Though the \textit{mask} feature sizes are equal for all designs, the \textit{filter radius} is larger for the designs with nonvertical sidewalls (\eqnref{eq:feature}). \textit{ii}) The field plot of the grating (\figref{fig:angled_evolution}d) shows that a significant amount of power is directed downward. This suggests that even though the etch angle breaks z-symmetry, the angled sidewalls are not advantageous towards minimizing substrate emission. Furthermore, the undercut etch (110$^\circ$) had significantly higher coupling than the shallow etch (70$^\circ$), though they had the same effective feature size (\eqnref{eq:feature}). This suggests that the undercut etch breaks z-symmetry in a way more useful than the shallow etch. However, the grating with vertical sidewalls significantly outperformed both. Therefore, we do not identify undercutting silicon as a viable option for increasing grating performance for these geometries.  Finally, we note that these designs were accomplished without geometric enhancements such as multiple/partial etches, buried mirrors, or exceptionally thick layers are commonly used to improve coupling efficiency \cite{cheng2020grating,chen2017grating}.

Enforcing a minimum feature size within a TO context is challenging for two reasons: \textit{i}) device performance is often highly dependent on small features and \textit{ii}) it is difficult to design a differentiable algorithm to determine the feature size of freeform geometries. Therefore, geometries often have features slightly smaller than the minimum feature size that has been enforced by constraints. This challenge is further exacerbated in TO for angled edges when a large filter is used to enforce a minimum feature size on the dilated geometry at the top of the layer. It is therefore practical to choose a filter size about $20\%$ larger than the theoretical feature size so the optimizer can discover a suitable local minimum early in the grayscale portion of the optimization without relying on small features close to the minimum feature size that will be challenging to expand or remove.


\section{3D design: LNOI dual-polarization s-bend}\label{sec:sbend}
\begin{figure}[ht!]
    \centering
    \includegraphics{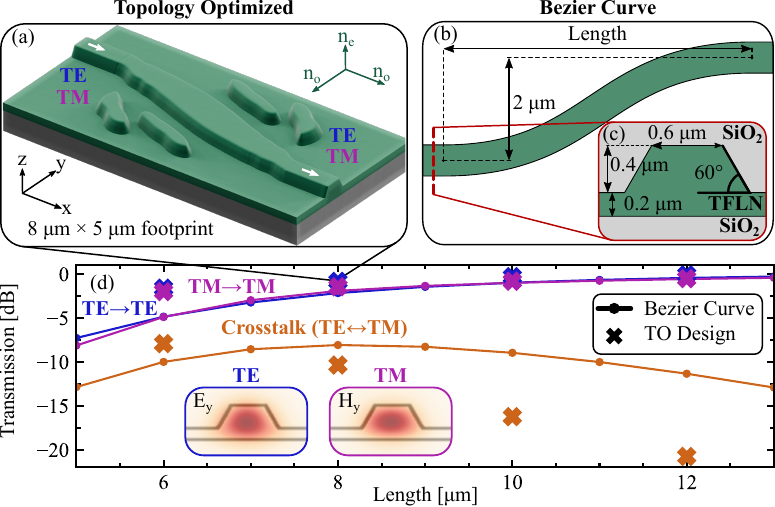}
    \caption{(a) An 8 $\upmu$m TO dual-polarization s-bend with superior transmission and polarization extinction to (b) the naive Bezier curve geometry. (c) Cross section of 600~nm lithium niobate layer with a 400~nm etch depth, etch angle of 60$^\circ$, and clad with SiO$_2$ (not shown in 3D render). (d) Simulated transmission and crosstalk for TO and Bezier curve s-bends across different lengths. The $E_y$ and $H_y$ mode fields for the TE$_{00}$ and the TM$_{00}$ mode respectively are shown in the inset.}
    \label{fig:s_bend}
\end{figure}

Dual-polarization s-bends are an essential component for high-throughput compact integrated photonic interconnects. S-bends with nonvertical etch angles, however, suffer from increased intermodal crosstalk between TE and TM modes compared to variants with vertical sidewalls\cite{sidewall_sbend}. The novel inverse-design methodologies enable compact s-bends with low polarization crosstalk on platforms that are difficult to etch vertically (e.g. LNOI) \cite{tfln_TO_sum,shang2023inverse,xue2025inverse,kim2025freeform}. Specifically,  we employ a 600~nm z-cut LNOI process with a 400~nm etch depth and 60$^\circ$ etch angle (\figref{fig:s_bend}c) to design s-bends with varying lengths that unilaterally outperform Bezier curve s-bends. 

This optimization uses a 2D implementation of the $\mathcal{S}$Morph transform given by \eqnref{eq:smorph}. We performed a parallel optimization to maximize the transmission of both the TE$_{00}$ and TM$_{00}$ modes (note the ``00'' subscript due to the 3D waveguides, in contrast with the 2D waveguides in Sec. 3 \& 4). The FOM is given by \eqnref{eq:FOM} where $\alpha_n^+$ is the overlap of the simulated fields in the n$^{\text{th}}$ mode (the same as the source mode) and $P_\text{n}$ is the power emitted by the source used to normalize the overlap integral. We optimized these objectives at four wavelengths evenly spaced between $\lambda=1530$~nm and $\lambda=1565$~nm which ensures broadband performance. The TO s-bends have a 2~$\upmu$m waveguide separation (center to center), a 6~$\upmu$m design region width, and a 6/8/10/12 $\upmu$m length (\figref{fig:s_bend}b). We applyed C$_2$ symmetry to the design variables which simplifies the design problem and equalizes the TM$\rightarrow$TE and TE$\rightarrow$TM crosstalk. During the final epoch of each design, geometric linewidth constraints were enforced to restrict the minimum feature size to 100~nm \cite{zhou_geometric}. The optimization ran until the constraints were satisfied and the performance showed minimal improvements. These devices were simulated with a resolution of 40 voxels/$\upmu$m and a design region resolution of 80 voxels/$\upmu$m \cite{hammond2022high}. To accomodate the high resolution 2D design parameters, we parallelized the 2D $\mathcal{S}$Morph transform across several cores. Distributed on three Intel Xeon Gold 6226 2.7 GHz nodes (totaling 72 cores) \cite{PACE}, each optimization ran for 48-72 hours for an average of $\mathbin{\sim}{100}$ iterations.

To evaluate the final performance of the s-bends, the design parameters were fully binarized with a Heaviside step function and simulated at 100 wavelength points from $\lambda$ = 1500~nm to $\lambda$ = 1600~nm. The 1-dB bandwidth was greater than the simulation band, indicating strong broadband performance. Compared to the Bezier curve-based designs, all TO designs had higher transmission through both polarizations and larger polarization extinction (\figref{fig:s_bend}d, Table \ref{tab:s_bend}). Bezier curve s-bends are commonly used in integrated photonics, often with sufficient lengths to avoid any significant intermodal coupling.

\begin{table}[ht]
    \centering
    \begin{tabular}{cccccc}
        \toprule
        {} & \multicolumn{2}{c}{Transmission} & {} & \multicolumn{2}{c}{Extinction}\\
        \cmidrule{2-3} \cmidrule{5-6}
        Length & TE & TM & Crosstalk & TE & TM  \\
        \midrule
        6 $\upmu$\text{m} & \textbf{$-$1.62}/$-$4.87 & \textbf{$-$2.01}/$-$4.90 & \textbf{$-$8.01}/$-$10.01 & \textbf{6.38}/5.14 & \textbf{6.04}/5.00 \\
        8 $\upmu$\text{m} & \textbf{$-$0.85}/$-$2.18 & \textbf{$-$1.46}/$-$1.93 & \textbf{$-$10.37}/$-$8.09 & \textbf{9.52}/5.91 & \textbf{9.02}/6.10 \\
        10 $\upmu$\text{m} & \textbf{$-$0.31}/$-$1.00 & \textbf{$-$0.83}/$-$1.02 & \textbf{$-$16.24}/$-$8.98 & \textbf{15.93}/7.98 & \textbf{15.72}/7.94 \\
        12 $\upmu$\text{m} & \textbf{$-$0.10}/$-$0.45 & \textbf{$-$0.53}/$-$0.58 & \textbf{$-$20.77}/$-$11.35 & \textbf{20.67}/10.90 & \textbf{20.28}/10.76 \\
        \bottomrule
    \end{tabular}
    \caption{Simulated transmission for both modes and crosstalk between modes for s-bends with four different lengths designed using either TO or Bezier curves ($\textbf{TO}$/Bezier).}
    \label{tab:s_bend}
\end{table}

The TO s-bends outperformed the Bezier s-bends in every metric. Both the transmission and extinction for both modes is higher for the TO s-bends compared with the Bezier s-bends of same length. All TO s-bends formed geometries similar to the naive s-bends, with an increased waveguide width in the center of the waveguide. Others have demonstrated Bezier curve s-bends with increased waveguide widths in the center that improve transmission \cite{BezierSi1}, but none have done this in LNOI. Side island features generated by the optimization for each TO device are not crucial to the operation of the device due to negligible field in these structures \cite{bobbyto}. The optimizer developed these features early in the optimization and were retained since they do not negatively impact the performance. Angled sidewalls and sharp bends are known to enhance intermodal coupling due to polarization rotation. These compact TO s-bends demonstrate that strategic optimization of the waveguide width throughout the bend can improve transmission and reduce intermodal crosstalk on a platform with angled sidewalls.

\section{Conclusion}\label{sec:conclusion}
We have developed novel TO methodologies for structures with nonvertical sidewalls and demonstrated them on angled and slanted etch profiles. The techniques are versatile, extending to a variety of material platforms including lithium niobate, and can be modified to support additional etch profiles such as positionally dependent angled sidewalls and blazed gratings by a combination of the shifted-$\delta$ and $\mathcal{S}$Morph transform. We showed that vertically stacking transformed parameters to create etch profiles can be accomplished using techniques consistent with density-based TO and associated features such as lengthscale constraints.

Future work includes using the sidewall mappings for more complex material stackups and design problems. For example, silicon nitride is successfully used in conjunction with LNOI to enable small feature sizes currently unaccessable by only etching the lithium niobate. Furthermore, we will develop formulations that produce devices robust to etch angle and optimize structures that commonly employ large bevel angles such as photodiodes. Lastly, \figref{fig:slanted_study} indicates that larger slant angles are preferable for chip-to-fiber coupling; moreover, jointly optimizing the slant angle as another degree of freedom alongside the design parameters can identify an ideal slant angle and potentially produce grating with higher coupling efficiencies.

\section*{Appendix A: Smooth grayscale erosions and dilations}
\label{sec:transforms}
From mathematical topology \cite{soille1999morphological}, the grayscale erosion $\ominus$ and dilation $\oplus$ operations for an image $f$ are defined as
\begin{equation}\label{eq:transforms}
    \begin{matrix}
        (f\ominus b)(x)=\inf\limits_{y\in E}\big\{f(y)-b(x-y)\big\} && (f\oplus b)(x)=\sup\limits_{y\in E}\big\{f(y)+b(x-y)\big\}
    \end{matrix}
\end{equation}
where $f$ is a 1,2, or 3D function $f:E\mapsto\mathbb{R}$ with $x\in E$ being the voxel coordinate and $b$ is called a structuring element. A structuring element a small set used to probe the image $f$ to transform it or extract information. The choice of structuring element often relies on apriori knowledge of the image and common examples include circles, pairs of points, or lines. In this work, a line was used for 1D erosions/dilations (Sec. 3,4) and circle was used for 2D erosions/dilations (Sec. 5).

The infimum and supremum operators (Eq. 16) preclude using gradient-based optimization techniques because they are inherently nondifferentiable. Therefore, to be used in a density-based TO framework, the grayscale erosion and dilation operators must be approximated by some smooth function. For this work, we use use the smooth morphology transform $\mathcal{S}$Morph, which is based on the $\alpha$-softmax function \cite{lange2014applications}. The $\alpha$-softmax, defined as:
\begin{equation}
    \mathcal{S}_\alpha=\cfrac{\sum_{i=1}^nx_ie^{\alpha x_i}}{\sum_{i=1}^ne^{\alpha x_i}}
\end{equation}
for some $\vect{x}=(x_1,...,x_n)\in\mathbb{R}^n$ and $\alpha\in\mathbb{R}$ approximates the min/max behavior of infinum/supremum operators but is differentiable. Importantly, $\lim\limits_{a\rightarrow\infty}\mathcal{S}_\alpha(\vect{x})=\max_ix_i$ and $\lim\limits_{\alpha\rightarrow-\infty}\mathcal{S}_\alpha(\vect{x})=\min_ix_i$. That is, the $\alpha$-softmax converges to the min and max operators when $|\alpha|$ becomes large. Similarly, the $\mathcal{S}$Morph operators
\begin{equation}\label{eq:smorph}
    \mathcal{S}\text{Morph}(f,w,\alpha)(x)=\frac{\displaystyle\sum\limits_{y\in W(x)}\big(f(y)+w(x-y)\big)e^{\alpha(f(y)+w(x-y))}}{\displaystyle\sum\limits_{y\in W(x)}e^{\alpha(f(y)+w(x-y))}},
\end{equation}
converge to the erosion and dilation operations
\begin{equation}
    \lim\limits_{\alpha\rightarrow+\infty}\mathcal{S}\text{Morph}(f,w,\alpha)(x)=\sup\limits_{y\in W(x)}\big\{f(y)+w(x-y)\big\}=(f\oplus w)(x)
\end{equation}
\begin{equation}
    \lim\limits_{\alpha\rightarrow-\infty}\mathcal{S}\text{Morph}(f,w,\alpha)(x)=\inf\limits_{y\in W(x)}\big\{f(y)+w(x-y)\big\}=(f\ominus -w)(x)
\end{equation}
as the parameter $|\alpha|$ becomes large. During the optimization phase, $|\alpha|=20$ was used to keep the optimization from stiffening and all final simulations in this work were performed with $|\alpha|=30$.

\section*{Appendix B: Thresholding approach for dilation and erosion}\label{sec:alternate}
We examine another method to perform erosions and dilations on the latent design variables. A common technique is to utilize the parameter $\eta$ which defined the center of the thresholding projection (Eq. 3) \cite{shang2023inverse,kim2025freeform}. An angled sidewall can be constructed using a height-dependent definition of $\eta$:
\begin{equation}
    \eta(z)=\eta_1+\frac{z-z_b}{z_t-z_b}\ \big(\eta_2-\eta_1\big)
\end{equation}
where $z_b$ and $z_t$ are the height of the bottom and top of the layer respectively and $\eta_1$ and $\eta_2$ are chosen enact the appropriate dilation on the top and bottom of the design layer. A significant challenge of this approach is that the width of the erosion/dilation is highly dependent on the topology of the parameters being eroded/dilated. Notably, this projection method requires the \textit{latent} design variables to be nearly binarized, a condition which is seldom met in TO. 

Figure 9 shows the sidewall angle projection for raw design variables that are not binarized. The result is variable sidewall angles in the projected field (Fig. 9a) as well as trenches that fill in at the bottom and ridges that completely erode at the top. This challenge is exacerbated by two scenarios: \textit{i}) when the erosion/dilation width is similar to or larger than the feature size and \textit{ii}) when the \textit{latent} design variables are nonbinary. While the linewidth constraints often impose sharp boundries in the latent design variables at the boundries of the thresholded design variables \cite{probst2024fabrication}, this requirement precludes grayscale optimization with thresholding-based erosions/dilations.

\begin{figure}[ht!]
    \centering
    \includegraphics{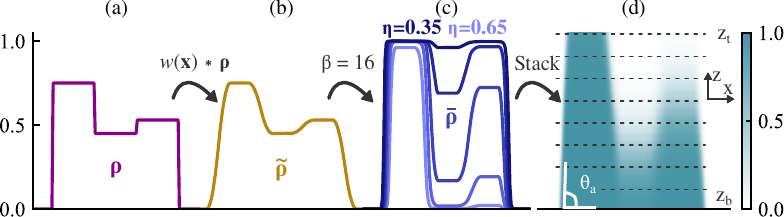}
    \caption{Design variables undergoing erosions/dilations by changing the center value, $\eta$ of the modified hyperbolic tanh projection (Eq. 3). The latent parameters (a) are filtered (b) and projected using values of $\beta=16$ and $\eta=0.35-0.65$. The thresholded parameters (c) are eroded or dilated differently depending on the slope of $\widetilde\vectrho$. Furthermore, they are pushed toward binarization for more significant amounts of erosions or dilations for the values of $\bar\vectrho$ that are not already binary.}
    \label{fig:eta_mapping}
\end{figure}

\begin{backmatter}
\bmsection{Funding}
This material is based upon work supported in part by the National Science Foundation (NSF) Center ``EPICA” under Grant No.1 2052808, \url{https://epica.research.gatech.edu/}. Any opinions, findings, and conclusions or recommendations expressed in this material are those of the author(s) and do not necessarily reflect the  views of the NSF. MJP and SER were supported by the Georgia Electronic Design Center of the Georgia Institute of Technology.

\bmsection{Acknowledgments}
The authors thank Alec M. Hammond for useful discussions. This research was supported in part through research cyberinfrastructure resources and services provided by the Partnership for an Advanced Computing Environment (PACE) at the Georgia Institute of Technology, Atlanta, Georgia, USA.
This material is based upon work supported in part by the National Science Foundation (NSF) Center ``EPICA'' under Grant No. 2052808, \url{https://epica.research.gatech.edu/}.

\bmsection{Disclosures}
The authors declare no conflicts of interest.

\bmsection{Data availability} Data underlying the results presented in this paper are not publicly available at this time but may be obtained from the authors upon reasonable request.

\end{backmatter}


\bibliography{references}

\end{document}